\documentclass[11pt, a4paper]{article}
\usepackage{jcappub, physunits}
\usepackage{setspace, caption, subcaption}
\usepackage{nccmath, physics}
\usepackage{xparse, mathtools}
\usepackage{aas_macros}
\usepackage[toc,page]{appendix}
\usepackage[mathscr]{euscript}
\newcommand{\crho}{\varrho}
\newcommand{\ep}{\epsilon}
\newcommand{\normal}{\mathcal{N}}
\newcommand{\bp}{\vb{p}}
\newcommand{\br}{\vb{r}}
\newcommand{\bv}{\vb{v}}
\newcommand{\bk}{\vb{k}}
\newcommand{\coll}{\mathcal{C}}
\newcommand{\mpl}{M_{\rm Pl}}

\title{Elastic Scattering of Cosmological Gravitational Wave Backgrounds: Primordial Black Holes and Stellar Objects}

\author[a,1]{Marcell Howard,\note{Corresponding author.}}
\author[b]{and Morgane K\"onig}

\affiliation[a]{Pittsburgh Particle Physics, Astrophysics, and Cosmology Center, Department of Physics and Astronomy, University of Pittsburgh, Pittsburgh, Pennsylvania, 15260, USA}
\affiliation[b]{Laboratory of Nuclear Science and Center for Theoretical Physics, Massachusetts Institute of Technology, Cambridge, Massachusetts, 02139, USA}

\emailAdd{mah455@pitt.edu}
\emailAdd{mkonig@mit.edu}

\abstract{Primordial black holes (PBHs) are plausible dark matter candidates that formed from the gravitational collapse of primordial density fluctuations. Current observational constraints allow asteroid-mass PBHs to account for all of the cosmological dark matter. We show that elastic scattering of a cosmological gravitational wave background, these black holes generate spectral distortions on the background of 0.3\% for cosmologically relevant frequencies without considering coherent scattering and 5\% when the coherent enhancement is included. Scattering from stellar objects induce much smaller distortions. Detectability of this signal depends on our ultimate understanding of the unperturbed background spectrum.}


\begin{document}
\maketitle

\section{Introduction}

The nature of dark matter (DM) is one of the most important questions still puzzling the physics communities. Numerous models have been proposed to resolve this problem. One such candidate, that has increasingly been receiving attention especially within the last decade, is primordial black holes (PBHs). PBHs are black holes that formed early in the Universe via gravitational collapse of primordial density fluctuations. Myriad of mechanisms have been identified that could have produced an appropriate spectrum of primordial density fluctuations, ranging from simple single-field models of inflation to multifield models with non-minimal couplings \cite{Carr20, Villanueva-Domingo21, Escriva22, Escriva_22, Chakraborty22, Geller22, Ozsoy23, Khlopov07, Belotsky14, Belotsky19, Ketov19, Bhattacharya23}. In addition to the vast literature on the production, formation, and abundance of PBHs, constraints on the fraction of DM they can constitute is just as varied \cite{Carr10, Carr16, Carr21, Green21}. These constraints span from Hawking evaporation and their subsequent $\gamma$-ray bursts on the low mass end to microlensing of stars in nearby galaxies (such as Andromeda) on the high-mass end. However, PBH masses within the range of $10^{-16}M_\odot \leq M \leq 10^{-11}M_\odot$ represent an entirely unconstrained region of parameter space and hence can still compose an $\order{1}$ fraction of DM. In this work, we show that gravitational waves (GWs) can be used to explore PBHs as DM candidates.

Indeed, GWs \cite{Abbott16, Abbott_16, Abbott17} are a promising avenue for investigating PBHs in general. The recent discovery of a stochastic gravitational wave background of either cosmological (isotropic) or astrophysical (anisotropic) origin by the North American Nanohertz Observatory for Gravitational Waves (NANOGrav) via the use of pulsar timing arrays (PTAs) \cite{Agazie23, Reardon23, Zic23, Antoniadis23a} expands the breadth of opportunities for scrutinizing the Universe over its lifetime. Previous studies on the constraining power of GWs have focused on calculating the merging rate of equal mass PBH binaries within the mass range of $10 - 300 M_\odot$ which could conceivably be detected by LIGO \cite{Ali-Haimoud17}. By assuming a monochromatic mass function, the DM fraction that can be attributed to PBHs was constrained to $\sim 10^{-3}$. More on the possible constraining capabilities of GWs as well as future forecasts can be found  here \cite{Sasaki18, Qin23}. In addition to these previous examinations, we propose to use the recent theoretical work done on the gravitational Sunyaev-Zeldovich (GSZ) effect \cite{Howard23}.

The GSZ effect is the gravitational analog of the more widely known SZ effect \cite{Sunyaev68} wherein distortions on the cosmic microwave background (CMB) are induced due to low energy Compton scattering of electrons occupying hot clouds of gas in the atmosphere of galaxy clusters \cite{Carlstrom02}. In our preceding manuscript \cite{Howard23}, we showed that Compton scattering of thermally produced beyond the standard model particles can generate similar distortions on any cosmological gravitational wave background (CGWB). 
We also provided estimates that showed that for frequencies that are sensitive to PTAs \cite{Moore15, Schmitz21}, can provide potentially large optical depths. In this article, we extend our analysis and show that this framework can be used to place additional constraints on PBHs by considering distortions on CGWBs due to Compton scattering. We also investigate the contribution due to stellar objects and stellar remnants, such as white dwarfs, as it is crucial to be able to understand the effect of various astrophysical objects on the CGWBs. We find that for most stellar object, the distortion on CGWBs is negligible except for red supergiants of mass $\order{10M_\odot}$.

The paper is organized as follows: in Section~\ref{pbhs}, we provide a review of the formalism of the GSZ effect where we extend it to include non-thermal DM candidates that follow a Maxwell-Boltzmann momentum distribution such as PBHs. In Section~\ref{stars}, we show how our new formalism can accommodate out of equilibrium objects such as stars and stellar remnants and in Section~\ref{discuss}, we conclude with some remarks, detection prospects, and directions for future work.

\paragraph{Conventions}

We use the mostly plus metric signature, i.e. $\eta_{\mu\nu} = (-,+,+,+)$, units where $c = \hbar = k_B = 1$, the reduced Planck mass $\mpl = (8\pi G)^{-1/2} \approx 2.43 \times 10^{18} \eV[G]$ and boldface letters $\br$ to indicate 3-vectors. Conventions for the curvature tensors, covariant and Lie derivatives are all taken from Carroll \cite{Carroll04} and all values for the cosmological parameters are taken from the Planck Collaboration 2018 results \cite{Planck20X}.

\section{Primordial Black Holes} \label{pbhs}

Our previous paper demonstrated that non-relativistic particles in a thermal bath will generate distortions on CGWBs due to Compton scattering. We can extend this analysis to incorporate non-thermal non-relativistic particles. The addition of these sorts of objects has the action of altering the momentum distribution from a simple thermal equilibrium Maxwell-Boltzmann distribution. PBHs are an example of out-of-equilibrium DM candidates where this would be relevant. In principle, if we start out with distribution function $\normal$, we would need $\normal = \normal(M, \bv)$, where $M$ and $\bv$ are the mass and velocity of the PBH. One would be inclined to posit a Maxwell-Boltzmann velocity distribution with a Press-Schechter mass distribution i.e.
\begin{equation}
    \normal(M,\bv) = \sigma(M_H)\exp(-\frac{\delta^2_c}{2\sigma^2(M_H)})\exp(-\frac{\bv^2}{2\sigma^2_v}),
\end{equation}
where $M_H$ is the mass of the horizon at the time the PBHs formed, $\delta_c$ is some critical or threshold overdensity, and $\sigma_v = 200\m[k]/\s$ is characteristic velocity of PBHs and stars moving in the Milky Way \cite{Binney87}. Our previous analysis can be carried over for these candidates so long as we can assign to them a characteristic energy $\ep$ so that we may have
\begin{equation}
    \Delta = \frac{\omega' - \omega}{T} = \frac{\omega}{T}\frac{\bp\cdot(\vu{n}' - \vu{n}) - \omega(1 - \vu{n}\cdot\vu{n}')}{E - \bp\cdot\vu{n}' + \omega(1 - \vu{n}\cdot\vu{n}')} \rightarrow \Delta = \frac{\omega' - \omega}{\ep} \equiv x' - x, \hspace{0.5cm}x = \frac{\omega}{\ep},
\end{equation}
where $\Delta$ is the dimensionless graviton energy shift, $\bp = M\bv$ is the co-moving incoming PBH 3-momentum, $\omega, \omega'$ are the incoming and outgoing graviton frequency with incoming and outgoing directions $\vu{n},\vu{n}'$ respectively and $x$ is the dimensionless frequency. Ref.~\cite{Oliveira21} gives us a prescription for assigning a characteristic energy and hence an effective temperature to a population that is out of thermal equilibrium. The closest thing to an effective temperature for PBHs is their Hawking temperature. At face value, given all the possible values for the mass of PBHs, one might object to using the Hawking temperature as a reliable effective temperature since it varies according to the mass of the black hole. One way to deal with it  is to consider a monochromatic mass function.\footnote{We are justified in using this approximation because the actual mass distribution tend to be strongly peaked around a typical or mean value of mass \cite{Niemeyer98, Green99, Kuhnel16, Young19, Biagetti21, Gow22}. The narrow width of the mass distribution enables us to approximate the distribution as a single delta function.} This approximation is particularly convenient for our purposes because the constraints on masses for which PBHs can account for an $\order{1}$ fraction of the total DM is narrow. Therefore we can fix the abundance of PBHs to be equivalent to cold DM (CDM), enabling us to assign their Hawking temperature as their characteristic energy $\ep = T_H \equiv (8\pi GM)^{-1}$. For an asteroid-mass PBH of $M = 10^{-16}M_\odot$, this corresponds to a Compton upscattering temperature of $T_H \sim \order{\eV[k]}$. While we lift the restriction of thermal equilibrium, the momentum distribution will still be the usual Maxwell-Boltzmann distribution\footnote{A Maxwell-Boltzmann distribution is sufficient for our purposes because PBHs can be taken to be a dilute collisionless non-relativistic gas such that the particles only interact via the Newtonian gravitational force \cite{Hektor18}.}
\begin{equation}
    \normal(\bp)\dd[3]{\bp} = \frac{n_{\rm PBH}}{(\sqrt{2\pi}M\sigma_v)^{3}}\exp(-\frac{\bp^2}{2M^2\sigma^2_v})\dd[3]{\bp},
\end{equation}
where $n_{\rm PBH}$ is the physical number density of PBHs given by $n_{\rm PBH}(z) = \rho_{\rm CDM}(z)/M = \Omega_{\rm CDM}\rho_c(1 + z)^3/M$ with $\Omega_{\rm CDM} = 0.2633$ and $\rho_c$ is the critical density today. Aside from the altered momentum distribution, the derivation of the non-thermal gravitational Kompaneets equation proceeds in much the same way as the thermal gravitational Kompaneets equation. Hence we refer the reader to Appendix \ref{lightning_grav_komp} for a lightning review and \cite{Howard23} for the complete treatment. We assume: $\omega \lesssim T_H$, soft gravitons ($\omega \ll M$), non-relativistic PBHs ($\abs{\bp} \ll M$), which implies a tiny energy shift $\Delta \ll 1$. We will therefore quote the result
\begin{equation}\label{grav_komp}
    \begin{split}
        \pdv{n}{t} &= A_T\qty[J_1(x,\lambda;s) + \frac{J_2(x,\lambda;s)}{2} + \frac{\alpha(J_1(x,\lambda;s) + J_2(x,\lambda;s))}{x} + \frac{\alpha(\alpha - 1)J_2(x,\lambda;s)}{2x^2}]x^\alpha,
    \end{split}
\end{equation}
where $n(x) = A_Tx^\alpha$ is the graviton occupation index\footnote{The occupation index is essentially the number of gravitons in state with frequency $\omega/T_H$.} evaluated at dimensionless frequency $x = \omega/T_H$ which generally follows a power-law with tensor amplitude $A_T = A_T(k_*)$ at pivot scale $k_* = 0.002\pc[M]^{-1}$. Deviations in the occupation index will be small so we plug the initial spectrum into the above equation. In general, we take $\alpha = -2$ as a representative case. Spin-$s$ dependent quantities, $J_\ell(x,\lambda;s)$, are given by
\begin{equation} \label{jell}
    J_{\ell}(x,\lambda;s) = 2\pi\int_{\theta_{\min}(\lambda)}^{\theta_{\max}}\sin\theta\dd{\theta}\int\dd[3]{\bp}\qty(1 - \frac{\bp}{m}\cdot\vu{n})\dv{\sigma_s(\bp,x,\theta)}{\theta}\normal(\bp)\Delta^{\ell}(x,\theta).
\end{equation}
These integral coefficients are to be understood as the average energy and energy squared that gets transferred from the PBHs to the gravitons. There is a $t$-channel pole due to the graviton exchange which is tied to the Coulombic nature of the gravitational potential. This makes it the same problem of the well known Rutherford scattering. Therefore we must regularize the $\theta$ integral by introducing cutoff angles $\theta_{\max}, \theta_{\min}(\lambda)$. We make the choice of cutoff by tying it to the geometric optics limit of linearized general relativity via a cutoff wavelength $\lambda$ as suggested in \cite{Cusin19}. That geometric optics limit occurs in a region $R_\lambda$ in which wave-like effects are non-negligible $R_\lambda = \qty(2\sqrt{3}r_s\lambda^2)^{1/3}$ where $r_s = 2GM$ and $\lambda = \lambda_{\rm GO}(1 + z)^{-1} $ is the wavelength of a typical gravitational wave and we need $r_s < \lambda$. This region gives a maximum value for the impact parameter $b_{\max} = R_\lambda$ and hence a minimum angle $\theta_{\min}(\lambda) = 2r_s/b_{\max}(\lambda)$. We also subsequently get a minimum value $b_{\min} = r_s$ for which absorption becomes relevant. Therefore there is also a maximum angle $\theta_{\max} = 2$. It should also be noted that for $M = 10^{-16}M_\odot$, we can safely assume that $r_s < \lambda_{\rm GO}$ which will not be the case when we consider extended objects like stars and stellar remnants.
\begin{figure}
  \centering
  \includegraphics[width=12cm]{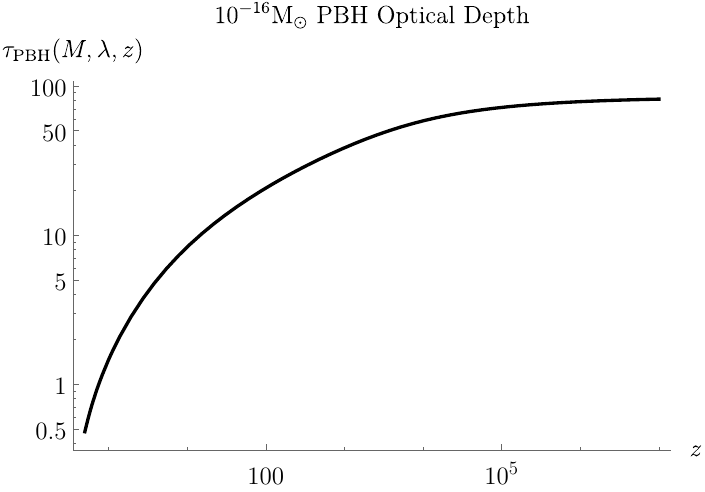}
  \caption{Optical depth of scattering for a CGWB with $10^{-16}M_\odot$ primordial black holes and cutoff wavelength $\lambda_{\rm GO} = 0.1\pc$ as a function of redshift.}
  \label{opt_depth}
\end{figure}
We can write down an explicit form for the integral coefficients $J_\ell$ for the spin-0 case
\begin{equation}
    \frac{\tilde{J}_1(x,\lambda;s = 0)}{x(4-x)} = \frac{\tilde{J}_2(x,\lambda;s = 0)}{2x^2} = -\frac{16\beta^3}{3} - 56\beta^2 - 112\beta - 256\log(\frac{1 - \beta}{2}) - \frac{172}{3},
\end{equation}
where $\beta = 1 - (2GM/\lambda)^{4/3}/2$ and $J_\ell(x,\lambda;s = 0) = n_{\rm PBH}\sigma_{\rm GC}(M, \lambda)\sigma_v^2\tilde{J}_\ell(x,\lambda;s = 0)$, $\sigma_{\rm GC}$ is the gravitational Compton cross section for graviton-scalar scattering \cite{Peters76}
\begin{equation}\label{cross_section}
	\begin{split}
		\sigma_{\rm GC}(M, \lambda) &= 2\pi(GM)^2\int_{\theta_{\min}(\lambda)}^{\theta_{\max}}\sin\theta\dd{\theta}\qty[\cot[4](\frac{\theta}{2})\cos[4](\frac{\theta}{2}) + \sin[4](\frac{\theta}{2})] \simeq \frac{4\pi(GM)^2}{q^2},
	\end{split}
\end{equation}
and $q = \sin((2GM/\lambda)^{2/3}/2) \approx (2GM/\lambda)^{2/3}/2$. Here we focus on slowly rotating PBHs since the angular momentum distribution for this particular range of masses is peaked around $s = 0$ \cite{DeLuca19, DeLuca20}, hence we will omit the explicit spin dependence. By recognizing that $\dv*{\tau_{\rm PBH}}{t} = n_{\rm PBH}\sigma_{\rm GC}$, we introduce the new Compton-$y$ parameter (i.e. the dimensionless energy transfer)
\begin{equation}
    y = \tau_{\rm PBH}(M, \lambda, z)\frac{\sigma^2_v}{(1 + z)^2},
\end{equation}
with $\tau_{\rm PBH}$ being the optical depth of scattering\footnote{The optical depth tells us how many gravitons scatter off a PBH in the time interval $[z_O, z_{\rm scat}]$. Thus an optical depth of $0.1$ means 1 out of every 10 gravitons will Compton scatter with a PBH at redshift $z_{\rm scat}$.}
\begin{equation}
    \begin{split}
        \tau_{\rm PBH}(z_{\rm scat}, z_O;M, \lambda) = \int^{z_{\rm scat}}_{z_O}n_{\rm PBH}(z)\sigma_{\rm GC}(M, \lambda, z)a(z)\dv{\eta}{z}\dd{z},
    \end{split}
\end{equation}
where $\eta$ is the conformal time and we have made the redshift dependence of the cross section more explicit. While it is common in the cosmological literature to write the optical depth as the first integral expression~\cite{Maggiore18}, we find the second integral expression to be a bit more useful for our purposes. We used an analytic expression for the scale factor that encompasses both the matter and radiation dominated eras $a(\eta)/a_{\rm eq} = 2\eta/\eta_* + \eta^2/\eta^2_*$ where $a_{\rm eq} = \Omega_R/\Omega_M$ is the scale factor at matter-radiation equality and $\eta_* = 2\Omega_R^{1/2}/H_0\Omega_M$. From here, one can simply invert this relation to get the conformal time as a function of the scale factor, and hence of the redshift. Because we are primarily interested in the matter and radiation dominated epochs, we fix the observed redshift $z_{O}$ to the matter-$\Lambda$ equality i.e. $1 + z_{O} = (\Omega_\Lambda/\Omega_M)^{1/3}$. Fig.~\ref{opt_depth} shows the optical depth as a function of redshift for our population of $10^{-16}M_\odot$ PBHs. We see that the optical depth gets larger at large redshift, reflective of the fact that a smaller universe guarantees the probability of scattering to be larger.
\begin{figure}
    \centering
    \includegraphics[width=12cm]{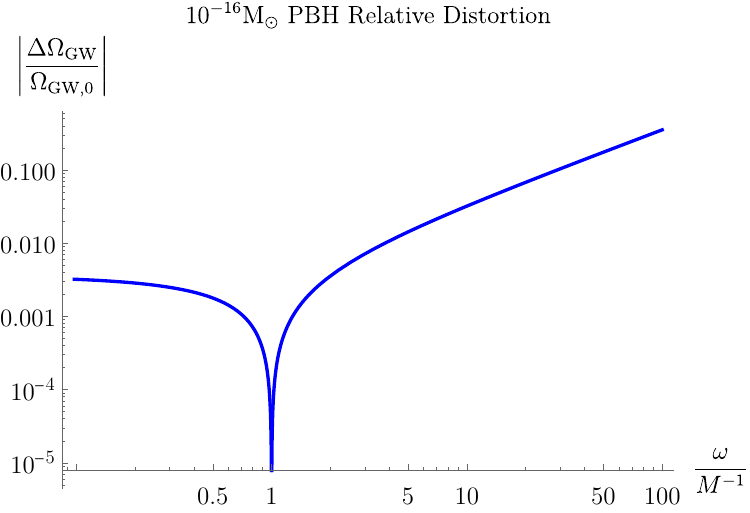}
    \caption{Absolute value of the relative distortion in the dimensionless energy density spectrum of a CGWB due to elastic scattering with $10^{-16}M_\odot$ PBH at a redshift of $z = 1.2$ with cutoff wavelength $\lambda_{\rm GO} = 0.1\pc$ in units of $8\pi G = 1$. The unperturbed spectrum is taken to be a power law spectrum for single-field inflation $\Omega_{\rm GW}\sim (\omega/M^{-1})^{2}$. The region $\omega/M^{-1} < 1$ corresponds to where $\Delta\Omega_{\rm GW}$ is negative.}
    \label{rel_dis}
\end{figure}
We can solve the gravitational Kompaneets equation by approximating $\pdv*{n}{y} \simeq \Delta n/y$
\begin{equation}
    \Delta n(x,z) = yA_T\qty[\tilde{J}_1(x,\lambda) + \frac{1}{2}\tilde{J}_2(x,\lambda) + \frac{\alpha(\tilde{J}_1(x,\lambda) + \tilde{J}_2(x,\lambda))}{x} + \frac{\alpha(\alpha - 1)\tilde{J}_2(x,\lambda)}{2x^2}]x^{\alpha},
\end{equation}
where $\tilde{J}_\ell(x,\lambda;s = 0) \equiv \tilde{J}_\ell(x,\lambda)$. Recall the dimensionless fractional energy density spectrum relates to the graviton occupation index by $\Omega_{\rm GW, 0}(x, z) = x^4n(x, z)$. Figure~\ref{rel_dis} shows the relative distortion in the energy density spectrum from single-field inflation with initial power-law spectrum $\Omega_{\rm GW} \sim x^{2}$ from elastic scattering with PBHs at redshift $z = 1.2$. We see that for $\omega/T_H < 1$, the spectrum is negative, which is indicative of a loss of power/energy at lower frequencies, and the distortion grows for $\omega/T_H > 1$. We can understand this from the analogous situation in the CMB\footnote{Unlike the CMB, the presence of an infrared divergence in this cross section causes lower frequencies tp get upscattered more than higher frequencies.} wherein because the particle number is conserved, the loss of power is a result of a loss of gravitons at lower frequencies. The interpretation of this phenomenon is simple: the Compton scattering of gravitons and PBHs results in a transfer of kinetic energy from the PBHs to the gravitons. The deficit in the background is due to a decrease of gravitons at low frequencies because they have been kicked up to higher energies leading to a corresponding surplus of gravitons at higher frequencies, resulting in an overall shifting of the spectrum. What's particularly interesting is that in the region $\omega < 0.1T_H$, the amplitude of the distorted spectrum is decreased by a factor of 0.3\%! We've neglected a factor that comes from the coherent scattering of PBHs. We show in Appendix \ref{coherent} that inclusion of this factor boosts the spectral distortion by an order of magnitude, leading to a spectral distortion of 5\% which is even more astounding.

\begin{figure}
    \centering
    \includegraphics[width=12cm]{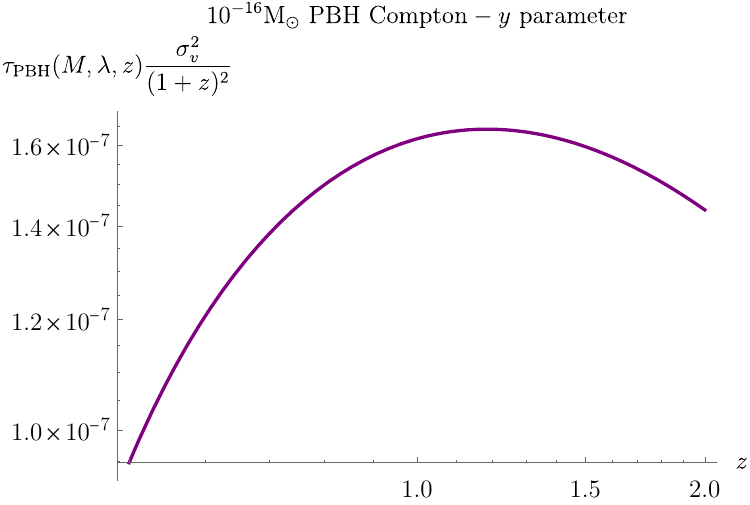}
    \caption{Compton-$y$ parameter for PBHs as a function of redshift, showing in which epoch to expect the greatest transfer of kinetic energy from the PBHs to gravitational waves. }
    \label{comp_y}
\end{figure}

Despite Fig.~\ref{opt_depth} showing that the optical depth grows to $\gg 1$, we chose $z = 1.2$ because it is approximately the redshift for which the transfer of kinetic energy from PBHs to the gravitons is greatest as seen in Fig.~\ref{comp_y}. Even though more gravitons will scatter with a PBH in a radiation-dominated Universe, those BHs do not have much kinetic energy and hence do not have much available energy to transfer to the gravitons. 
It is important to stress that the actual measurable effect occurs in the region $\omega < 0.1 T_H$ and our formalism is only applicable for $\omega < 10T_H$. However, we give some evidence of the validity of our results in the region of $T_H \leq \omega \leq 10T_H$ in Appendix \ref{long_wave}. The frequencies at which Compton upscattering takes place are associated with \eV[k] temperatures. This is evident from the curve of the relative distortions in the energy density spectrum, which turns over at approximately $\omega \simeq T_H$ corresponding to $\omega \sim 10^{18}\Hz$, aligning it with X-rays in the electromagnetic spectrum.

These frequencies lie within the realm of future tabletop GW detectors (see Ref.~\cite{Aggarwal21} for a recent review of proposed future experiments). For frequencies that are presently available to us \cite{Moore15, Robson19, Schmitz21} PBHs with mass $10^{-16}M_\odot$ and redshift $z = 1.2$, gives rise to a decrease in power that corresponds to $10^{-3}$ of the original amplitude of the background. It must also be stated that these estimates represent the most optimistic scenario for placing constraints on PBHs and therefore not necessarily what one would immediately measure.

\begin{table}[t]
    \centering
    \begin{tabular}{|c|c|c|c|c|}
    \hline
     Star Type & $\crho = R/R_\odot$ & $\mu = M/M_\odot$ & $\Delta\Omega_{*}(b < R)/\Delta\Omega_{\rm PBH}$ & $\Delta\Omega_{*}(b > R)/\Delta\Omega_{\rm PBH}$ \\
     \hline
     Main Sequence & 1 & 1 & $10^{-6}$ & $10^{-8}$ \\
     \hline 
     White Dwarfs & 0.01 & 0.5 & $10^{-9}$ & $10^{-15}$ \\
     \hline 
     Red Supergiants & 900 & 20 & 0.1 & $10^{-9}$ \\
     \hline
    \end{tabular}
    \caption{Radii, masses, and ratio of distortion from stars and PBHs for when GWs fall inside and outside the object's radius. Estimates of Betelgeuse were taken to be representative of red supergiants. Assumed white dwarf and red supergiant fraction was 5\% and 1\% respectively.}
    \label{star_table}
\end{table}

\section{Stars and Stellar Remnants} \label{stars}

Our formalism can be adapted to include stars on the main sequence as well as stellar remnants like white dwarfs. Since the average mass and radius for these populations are different, it is sensible to study the effects of elastic scattering of these two populations separately. Gravitational Compton scattering with any kind of extended object broadly fit within two regimes: the impact parameter, $b$, is $b > R$ where $R$ is the radius of the extended object and $b < R$. $10^{-16}M_\odot$ PBHs fit into the $b > R$ regime.\footnote{Asteroid PBH have a Schwarzschild radius of $10^{-12}\m$ making their effective size comparable to atoms.} \cite{Cusin19} gives a nice prescription for dealing with both cases which we will quote the estimate for the optical depth
\begin{equation}
    \tau_*(z) = \frac{n_*}{H_0\Omega^{1/2}_M}\begin{cases}
        \frac{2}{3}\sigma_{\rm ext}[(1 + z)^{3/2} - 1], & b < R\\ 6\sigma_{\rm GC}[(1 + z)^{1/6} - 1], & b > R
    \end{cases}
\end{equation}
where $n_*$ is the co-moving number density of stars in the universe given by $n_* = 0.1\rho_B/M_\odot = 0.1\Omega_B\rho_{c}/M_\odot \simeq 0.1\pc[k]^{-3}$ with $\Omega_B = 0.0492$, $\sigma_{\rm ext}$ is the gravitational Compton cross section for an extended object given by
\begin{equation}
    \sigma_{\rm ext}(\crho, \mu, b) = 10^{-5}\frac{\crho^4}{\mu^2}\qty(\frac{b}{R})^{10} \pc^2,
\end{equation}
with $R = \crho R_\odot$ and $M = \mu M_\odot$. For a uniformly distributed impact parameter $b \in [0, R_\odot]$, we use the mean as our estimate $\expval{b} = R_\odot/2$. Table~\ref{star_table} displays the radii, masses, $\sigma_{\rm ext}$, $\sigma_{\rm GC}$, and suppression factor relative to the relative distortion from PBH scattering found in fig.~\ref{rel_dis}.

The curve of the relative distortions on the background due to scattering from stars essentially acts as a re-scaling of the distortion generated by PBHs. We find a generic suppression for all objects. This is due to the fact that the optical depth scales like $M^{-1/3}$ for masses above the Planck scale.\footnote{For masses below the Planck scale, the optical depth scales like $\tau \sim M^{14/3}$ \cite{Howard23}.} It would thus require an extremely precise measurement of the CGWB to one part in $10^{11}$ in order to detect distortions in the background from Compton scattering of stars. A prospect that seems quite far off from current technological capabilities. The problem is made worse when considering white dwarfs. Their relatively small cross section paired with their tiny fraction of the overall number of stars makes it so the induced distortions is even tinier. Red supergiants seem to be the only stellar candidates with an effect that could feasibly be detected. However, the average mass and radius for these objects are highly model dependent, therefore it is difficult to make any definitive statements on this population.

\section{Discussion and Future Outlook} \label{discuss}

In summary, we have demonstrated that the gravitational SZ (GSZ) effect for particles in thermal equilibrium can be easily extended to include objects that are not in thermal equilibrium but are still collisionless and only interact through a Newtonian potential. In doing so, we have shown that distortions due to elastic scattering between gravitons and primordial black holes, stars and stellar remnants can be expected to be found on any CGWB. For $10^{-16}M_\odot$ PBHs, we can place constraints on the abundance of these DM candidates when considering the low frequency bands of PTAs. It was then shown that elastic scattering generates a broken power law on the initial background spectrum wherein at low frequencies, we get a 0.3\% decrease in the background's amplitude when the enhancement from coherent scattering is left out. In Appendix \ref{coherent}, we demonstrate that including this additional coherent enhancement factor gives us a factor of 10 in the distortion, bringing the spectral distortions to about 5\%. Lastly, we have demonstrated that while distortions made by stars and white dwarfs will be too minor to be detectable for the foreseeable future. However, red supergiant stars might produce observable signals in any CGWB.

We would like to end this paper on a discussion of measurement prospects. In order to extract this signal, there are two considerations that must be confronted: are the current experiments' sensitivities sufficient to probe the potential signal and is the unperturbed background spectrum truly understood? The former concern is well beyond the scope of this manuscript and is an issue we leave in the more capable hands of instrumentalists. We shall discuss the latter in some detail.

In the case of the CMB, attributing different distortions to particular physical phenomena is comparatively easy because we know that the initial spectrum is a near-perfect black body. The same can not be said for CGWBs\footnote{However, for redshifts $z\sim 1$, one could observe spectral distortions of the anisotropies e.g. \cite{Alba16, Contaldi17, Bartolo19} to identify the presence of Compton scattering. Thank you to Benjamin Lehmann for pointing this out.} because the unperturbed spectra can only be approximated as certain power laws in different regimes. Detection of CGWBs is already made complicated by the presence of astrophysical foregrounds \cite{Biscovenau20, Zhou22a, Zhou22b} as they are expected to dominate over the signal of any CGWB. 

An inflationary background would provide the cleanest signal for these distortions. It's nearly scale invariant behavior at all frequencies makes the presence of a broken power law particularly evident. However, in addition to the expected amplitude of the inflationary background being small \cite{Maggiore00}, it has been shown that at approximately 25 e-folds before the end of slow roll,\footnote{This epoch of inflation would produce GWs that could be detected by LISA or PTA experiments.} inflation ceases to be the simple $\Omega_{\rm GW} \sim \omega^{2}$ power law \cite{Caligiuri15} from single-field inflation. These additional features would need to be well constrained first before any serious claim for detecting the effect of PBH elastic scattering on the background can be made.

We can also consider a background from first order phase transitions in the early universe \cite{Kamionkowski94, Gogoberidze07, RoperPol20}. The amplitude of the background is much larger compared to inflation $\Omega_{\rm GW} \sim 10^{-12}-10^{-14}$ and the dimensionless energy density spectrum is a power law $\Omega_{\rm GW} \sim (\omega/T_H)^{2.8}$ at low frequencies with a characteristic peak frequency followed by a decaying power law $\Omega_{\rm GW} \sim (\omega/T_H)^{-1.8}$ but whose peaks are dependent on either the amount of turbulence of the colliding bubbles or the amount of acoustic waves which are generated in the cosmic plasma. Therefore a detection of a distortion of $\sim 0.3\%$ would be difficult in the region of the peaks since the exact details of the role of turbulence is currently an active field of research. The best case scenario would be for the Compton-upscatter to occur at low frequencies since the signal would be much cleaner. However, given that we are considering PBHs with a Hawking temperature of $T_H \sim \order{\eV[k]}$, this prospect seems unlikely.

The main limitation of our study is the use of a cutoff for the gravitational Compton cross section. From the perspective of fundamental physics, a cutoff is unsavory for at least two reasons: (1) Lorentz and gauge invariance is loss due to the artificially restricted phase space and (2) our final answer depends on the cutoff. As a result, there is a question as to how large of an error is introduced from the cutoff. Another drawback of our chosen infrared cutoff is that we have assumed that it is valid for all frequencies indicated in Fig.~\ref{rel_dis} which should be further examined in future works. 

A potentially interesting direction would be considering the role angular momentum plays in generating distortions on any CGWB. Previously, we showed that induced distortions were proportional to the quantum mechanical spin of the massive scatterer. While we only focused on PBHs that typically have no angular momenta, we know that most conventional black holes are spinning. Thus, even if the optical depth scales like $M^{-1/3}$, it is feasible that a decrease in the mass can be compensated by considering a population of near-extremal black holes. However, one would need to go back and derive a new expression for $R_\lambda$, i.e. the region in which the geometric optics approximation for gravitational scattering holds. In this case, the Kretschmann scalar will correspond to the Kerr-metric of a spinning black hole. 

Lastly, while our formalism was entirely focused on a CGWB, adapting this calculation technology to an astrophysical background is quite feasible. The main difference is the presence of additional spatial derivatives in the Boltzmann equation and some work has already been done on this front \cite{Pizzuti22}. If the NANOGrav background is resolutely determined to be an astrophysical background of binary black holes, one could study the effects that elastic scattering from $10^{-16}M_\odot$ PBHs would induce that background as well. Since Fig.~\ref{comp_y} shows that the dominant signal is coming from $z \simeq 1.2$, this distortion would be a worthwhile venture to pursue.

\section*{Acknowledgements}

The authors would like to thank Jordan Scharnhorst, Arthur Kosowsky, David Kaiser, and Andrew Zentner for reviewing the first draft of this manuscript as well as providing useful comments to bolster this article. The authors also thanks Tatsuya Daniel for his initial participation in the project. M.H. would like to especially thank Arthur Kosowsky for directing the authors' attention toward PBHs as a potentially interesting candidate to study with the GSZ effect. M.K. would also like to thank Aya Milan K\"{o}nig Siebke for her patience. Mathematica plots were made with \emph{MaTeX} \cite{Horvat22} and we also made use of the helpful unit conversion tables found in \cite{Tomberg21}. This project has been partially completed at the Laboratory for Nuclear Science and the Center for Theoretical Physics at the Massachusetts Institute of Technology (MIT-CTP/5620).

\appendix

\section{The Gravitational Kompaneets Equation: Lightning Derivation}\label{lightning_grav_komp}

We give a brief summary of the derivation of Eqn. (\ref{grav_komp}) for which one could look at Ref. \cite{Howard23} for the full details. We start from the assumption that dark matter initially occupied a thermal bath at temperature $T$ with mass $m$. We assume the following: minimal coupling to gravity, $|\bp|$ and $T\ll m$ i.e. the dark matter is non-relativistic, and soft gravitons i.e. $\omega \ll m$ where $\omega$ is the energy of the graviton but $\omega\sim T$. These assumptions combined leads to the energy shift of the gravitons i.e.
\begin{equation}
    \Delta = \frac{\omega' - \omega}{T} = \frac{\omega}{T}\frac{\bp\cdot(\vu{n}' - \vu{n}) - \omega(1 - \vu{n}\cdot\vu{n}')}{E - \bp\cdot\vu{n}' + \omega(1 - \vu{n}\cdot\vu{n}')},
\end{equation}
to be small i.e. $\Delta \ll 1$. Since we are interested in the statistical evolution of the gravitons, we start with the Boltzmann equation
\begin{equation}
    \pdv{n(\bk,t)}{t} = \coll[n],
\end{equation}
where $n(\bk,t)$ is the graviton occupation index/number with momentum $\bk$ and $\coll[n]$ is the collision term given by
\begin{equation}
    \begin{split}
        \coll[n] = \int\dd[3]{\bp}\int\dd[3]{\bp'}\int\dd[3]{\bk'}&\left[\normal(\bp')n(\bk',t)w(p',k'\rightarrow p,k)(1 + n(\bk,t))\right.\\ &-\left. \normal(\bp)n(\bk,t)w(p,k\rightarrow p',k')\qty(1 + n(\bk',t))\right].
    \end{split}
\end{equation}
Here $\normal(\bp)$ is the Maxwell-Boltzmann momentum distribution for fundamental particle dark matter in thermal equilibrium
\begin{equation}
    \normal(\bp) = \normal_{\rm eq}(|\bp|) = \frac{n_{\rm CDM}}{(2\pi mT)^{3/2}}\exp(-\frac{\bp^2}{2mT}),
\end{equation}
with $n_{\rm CDM}(z) = \Omega_{\rm CDM}\rho_c(1 + z)^3/m$ is the physical number density. We've also introduced $w(p',k'\rightarrow p,k)$ as the non-covariant transition rate
\begin{equation}
    w(p,k\rightarrow p',k')\dd[3]{\bp'}\dd[3]{\bk'} = \qty(1 - \bv\cdot\vu{n})\dv{\sigma_s(\bp,\bk,\bk')}{\Omega}\dd{\Omega},
\end{equation}
with $(1 - \bv\cdot\vu{n})$ being the M$\o$ller velocity and $\dv*{\sigma_s}{\Omega}$ is the differential gravitational Compton cross section for an arbitrary spin-$s$ massive target. We can make a change of variables $\omega\rightarrow x = \omega/T$, $\omega'\rightarrow x' = \omega'/T$, and expand $\Delta$ to quadratic order i.e. write $x' = x + \Delta$ and Taylor expand the integrand with $n(x',t) = n(x + \Delta,t)$ which leads to the following equation
\begin{equation}
    \pdv{n}{t} = \frac{J_2(x,\lambda;s)}{2}\pdv{x}\qty[\pdv{n}{x} + n(1 + n)] + \qty(J_1(x,\lambda;s) + \frac{J_2(x,\lambda;s)}{2})\qty[\pdv{n}{x} + n(1 + n)].
\end{equation}
Lastly since the occupation index for gravitons follows a power law $n(x) = A_Tx^\alpha$ and we expect the distortions to be small, we insert this into the above expression while neglecting terms $\order{A_T^2}$ and we get Eqn. (\ref{grav_komp}).

\section{Coherent Scattering Limit of Primordial Black Holes}\label{coherent}

In our original formalism for describing scattering of gravitational wave backgrounds, we assumed single particle scattering. If PBHs with masses $M = 10^{-16}M_\odot$ are assumed to be dark matter, we can have a coherent enhancement from the GW "seeing" a cluster of PBHs as opposed to singular BHs. Let $N$ be the number of PBHs that falls within the wavelength of the GW. Thus, the mass increases by $M\rightarrow NM$ which implies the cross section is enhanced by a factor of $\sigma_{\rm GC} \rightarrow N^{2/3}\sigma_{\rm GC}$. Let us put in some numbers to figure out how this enhancement figure will affect our result. The average spacing, $\ell$, of these PBHs can be shown to be
\begin{equation}
    \ell = \qty(\frac{M}{\rho_{\rm CDM}})^{1/3} \simeq 10^{-3}\pc.
\end{equation}
However, the wavelength that we've taken our gravitational wave detector to be sensitive up to is $\lambda_{\rm GO} = 0.1\pc$. This means that the coherent enhancement factor is
\begin{equation}
    N = \frac{\lambda_{\rm GO}}{\ell} \simeq 70.
\end{equation}
This means that the cross section gets enhanced by a factor of $10^{4/3}$. Since the GSZ effect scales linearly with the cross section, the spectral distortion at PTA frequencies goes from 0.3\% to 5\%!

\section{Beyond the Long Wavelength Approximation}\label{long_wave}

Our analysis has primarily focused on the long wavelength regime of gravitational scattering i.e. $M\omega \ll 1$ where we set $8\pi G = 1$. However, there is some question about the regime where $T_H < \omega < 10T_H$. Here $\omega$ is the initial frequency of the graviton. We are particularly interested in answering how large do the corrections to the cross section for gravitational scattering become? First, we write down the differential cross section in the rest frame of the massive particle
\begin{equation}
    \dv{\sigma_{\rm GC}}{\Omega} = M^2\qty(\frac{\omega'}{\omega})^2\qty[\cot[4](\frac{\theta}{2})\cos[4](\frac{\theta}{2}) + \sin[4](\frac{\theta}{2})],
\end{equation}
where $\omega'/\omega$ is a phase-space factor equal to
\begin{equation}
    \frac{\omega'}{\omega} = \frac{1}{1 + \frac{2\omega}{M}\sin[2](\frac{\theta}{2})},
\end{equation}
and in the long wavelength approximation, $(\omega'/\omega)^2 \simeq 1$. Taking a step beyond the long wavelength regime corresponds to adding additional terms from i.e.
\begin{equation}
    M^2\qty(\frac{\omega'}{\omega})^2 = M^2 - 4M\omega\sin[2](\frac{\theta}{2}) + \ldots
\end{equation}
We still have an infrared divergence so we need to implement a cutoff i.e.
\begin{equation}
    \sigma_{\rm GC} = \int_{\theta_{\min}(\lambda)}^{\theta_{\max}}\dd{\Omega}\dv{\sigma_{\rm GC}}{\Omega}.
\end{equation}
Plugging in the leading order correction to the phase-space factor into the differential cross section, while also maintaining the values we plugged in for $M = 10^{-16}M_\odot$ and $\lambda = 0.1\pc$, we find that the dominant contribution to the cross section is still $q = (2GM/\lambda)^{2/3}/2$ i.e. the cross section still looks like the one we attained in Eqn. (\ref{cross_section}).

\bibliographystyle{JHEP}
\bibliography{primordial_black_hole_compton_scattering}

\end{document}